\documentclass[%
 reprint,
superscriptaddress,
 amsmath,amssymb,
 aps,
]{revtex4-2}

\usepackage{graphicx}
\usepackage{dcolumn}
\usepackage{bm}

\begin{document}

\preprint{APS/123-QED}

\title{Isotope Shifts in the Metastable $\mathbf{a^5F}$ and Excited $\mathbf{y^5G^\circ}$ Terms of Atomic Titanium}

\author{Andrew O. Neely}
 \email{neelya@berkeley.edu}
  \affiliation{Department of Physics, University of California, Berkeley, Berkeley, CA 94720, USA}
 \affiliation{Challenge Institute for Quantum Computation, University of California, Berkeley, CA 94709, USA}
 \author{Kayleigh Cassella}
 \affiliation{Department of Physics, University of California, Berkeley, Berkeley, CA 94720, USA}
 \affiliation{Atom Computing Inc., Berkeley, CA 94710, USA}
 \author{Scott Eustice}
  \affiliation{Department of Physics, University of California, Berkeley, Berkeley, CA 94720, USA}
 \affiliation{Challenge Institute for Quantum Computation, University of California, Berkeley, CA 94709, USA}
 \author{Dan M.\ Stamper-Kurn}
 \email{dmsk@berkeley.edu}
 \affiliation{Department of Physics, University of California, Berkeley, Berkeley, CA 94720, USA}
 \affiliation{Challenge Institute for Quantum Computation, University of California, Berkeley, CA 94709, USA}
 \affiliation{Materials Science Division, Lawrence Berkeley National Lab, Berkeley, CA 94720, USA}

\date{\today}

\begin{abstract}
%Atomic isotope shifts evince a variety of physical phenomena, owing chiefly to the involvement of both electronic and nuclear structure considerations. Since they straddle the interface between nuclear and electronic measurements, investigations involving atomic isotope shifts are uniquely poised to probe effects involving the structure of the atom at all scales, as well as the interplay between these effects. Titanium's isotope shifts are a particularly apt system for studying an assortment of facets of atomic structure, including changes in electron density, electronic correlation energy, and even light bosonic fields. Here, we measure and analyze the isotope shifts of the five $J\to J+1$ and four $J\to J$ transitions between the metastable a$^5$F and excited y$^5$G$^\circ$ energy terms in titanium for its three stable bosonic isotopes ($^{46}$Ti, $^{48}$Ti, and $^{50}$Ti) using saturated absorption spectroscopy in a hollow cathode lamp. Our results begin to yield insight into the electronic and nuclear structure of titanium and enable future explorations in atomic structure theory.
We measure and analyze the isotope shifts the multiplet of transitions between the metastable a$^5$F and excited y$^5$G$^\circ$ terms of neutral titanium by probing a titanium vapor in a hollow cathode lamp using saturated absorption spectroscopy. We resolve the five $J\to J+1$ and the four $J\to J$ transitions within the multiplet for each of the the three $I=0$ stable isotopes ($^{46}$Ti, $^{48}$Ti, and $^{50}$Ti). The isotope shifts on these transitions allow us to determine the isotope-dependent variation in the fine-structure splitting of the a$^5$F and y$^5$G$^\circ$ levels themselves. Combined with existing knowledge of the nuclear charge radii of titanium nuclei, we derive the specific mass and field shifts, which arise from correlated electronic motion and electronic density at the nucleus respectively, and further observe a strong $J$-dependent variation in each. Our results yield insight into the electronic and nuclear structure of transition metal atoms like titanium, and also characterize optical transitions that may allow for optical manipulation of ultracold gases of transition metal species.\end{abstract}

%found these bits in the abstract kinda floating around
%We ascribe the energy shifWe observe a strong $J$-dependent variation in the between the metastable a$^5$F and excited y$^5$G$^\circ$ energy terms in titanium for its three stable bosonic isotopes ($^{46}$Ti, $^{48}$Ti, and $^{50}$Ti) using saturated absorption spectroscopy in a hollow cathode lamp. Our results begin to yield insight into the electronic and nuclear structure of titanium and enable future explorations in atomic structure theory.

\maketitle

The isotope shifts of atomic spectra encode information about the nuclear and electronic structure of the atom \cite{king1}. %\cite{king1,king2,titaniumshifts,titaniumshifts2,titaniumshifts3,titaniumshifts4,hfs_is_uv,sms,sms2,iron}.
The dominant sources of these shifts are the normal mass shift, which occurs from a change in the system's reduced mass, the specific mass shift, which reflects correlations within the electronic wavefunction, and the field shift, which stems from changes in nuclear volume. As such, precise isotope shift measurements provide strong tests of atomic and nuclear structure theory \cite{correlationEnergy,density1,density2,density3, calcium,indium}. A more precise examination of isotope shifts is also proposed as a means to reveal electron-nucleus interactions beyond those within the Standard Model \cite{is-bsm1,is-bsm2}. Aside from probing atomic structure, isotope shifts also provide necessary information for isotope-specific optical manipulation of atoms and the interpretation of astrophysical spectra \cite{cleg79isotope,pavl20isotope}.

% provides insight into the interplay between nuclear size and various electronic parameters, including density, correlation energy, and changes to the reduced mass. For these reasons, the IS of atomic spectra is a particularly apt testing ground for various theoretical endeavours, ranging from atomic structure theory  to physics beyond the standard model, where IS have been postulated as a sensitive probe of light bosonic fields \cite{is-bsm1,is-bsm2}.

Transition metal atoms provide compelling targets for isotope-shift investigations because of their intermediate complexity. Compared with alkali and alkali-earth atoms, transition metals provide the additional richness of having partially filled d-subshells. These generate multiple same-parity atomic states that are admixed by electron-electron interactions so as to yield large electron-electron correlations, which can be probed via the specific mass shift. The d-subshell also shields the nuclear charge, thereby influencing the electronic charge density at the nucleus, and therefore is evident in measurements of the field shift. At the same time, transition metals are less complex than heavier elements such as lanthanides and actinides, and are thus more amenable to theoretical descriptions.

Among the transition metals, titanium is a particularly attractive choice for such isotopic characterization \cite{titaniumbosons,titaniumshifts,titaniumshifts2,titaniumshifts3,titaniumshifts4,tij,anastassov_optical_1994, hfs_is_uv}. Titanium's smaller atomic number and large isotopic selection make it a good model for detailed atomic theory, compared to heavier transition metals, most notably due to the lower relativistic effects. Titanium's four valence electrons provide a rich complexity that is still accessible by theoretical techniques, and the near-degeneracy of the 3$d$ and 4$s$ levels adds to this \cite{tij}. Titanium has also been targeted for its tightly packed atomic level structure, which leads to many states marked by significant state mixing \cite{tij,hfs_is_uv}. Titanium has also played an important role in nuclear physics, because its  proximity to the magic proton number of $Z=20$ and $^{50}$Ti's magic neutron number of $N=28$ results in a trend of decreasing nuclear size for increasing nucleon number \cite{nuclear,nuclear2,hfs_is_uv}. Additionally, atomic titanium has recently been identified as a candidate for laser cooling \cite{transitioncooling}. Isotope shift investigations may, therefore, provide key information toward generating a new family of quantum degenerate atomic gases.

\begin{figure}[htb]
    \includegraphics[width=0.9 \columnwidth]{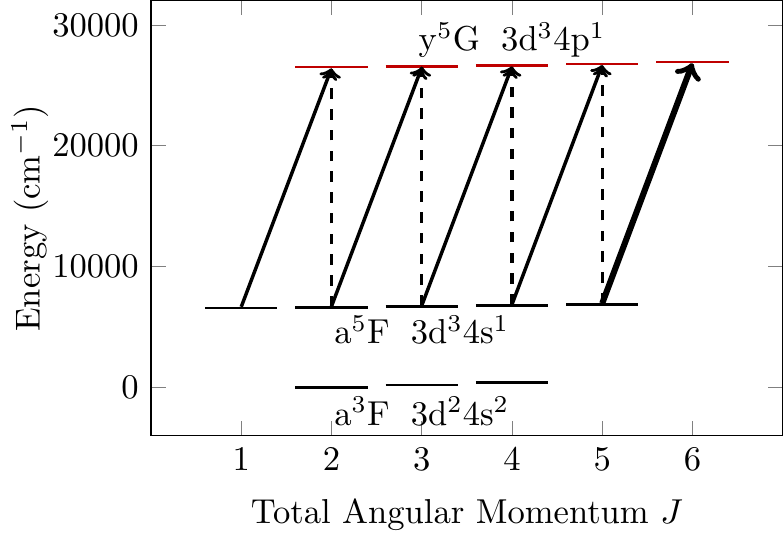}
    \caption{The relevant energy levels in titanium. The $a^5F_5$ metastable state supports a near-cycling transition (bolded) to the $y^5G^\circ_6$ state. The $J\to J+1$ transitions (solid) and $J\to J$ transitions (dashed) together can be used to determine the level shifts of all of the levels in the a$^5$F$_J$ and y$^5$G$^\circ_J$ manifolds. Black levels are even-parity, and red (upper) levels are odd-parity.}
    \label{fig:levels}
\end{figure}

Here, we apply Doppler-free spectroscopy to resolve optical transitions of the three zero-nuclear-spin isotopes of titanium ($^{46}$Ti, $^{48}$Ti and $^{50}$Ti; see Table \ref{tab:isotopes}) and  measure the isotope shifts on the multiplet of transitions between the a$^5 $F and y$^5$G$^\circ$ terms of atomic titanium, as shown in Fig. \ref{fig:levels}. The measured shifts on both the $J \rightarrow J$ and $J \rightarrow J+1$ transitions allow us to determine the isotope shift and its $J$-dependence of the fine structure splitting of both the a$^5$F and y$^5$G$^\circ$ terms. Similar studies have focused on $J$-dependent isotope shift, noting effects from three distinct sources: spin-other-orbit (SOO), relativistic, and crossed-second-order (CSO) interactions \cite{tij, hfs_is_uv, bauche}. The terms studied here are largely free of admixture and therefore from the complication for CSO terms in the isotope shift, allowing for a cleaner measurement of $J$-dependence due to SOO and relativistic interactions. These two levels are further attractive for measurements of the $J$-dependence of isotope shifts because they support a larger range of $J$ than similar measurements in the past \cite{tij, hfs_is_uv}.

We observe clear $J$-dependent variations of the isotope shift, and interpret these variations as reflecting the $J$-dependencies of electronic correlations and electron density in titanium. Among the measured transitions is the a$^5$F$_5\!\rightarrow$y$^5$G$_6^\circ$ transition that has been identified as a near-cycling transition suited for laser cooling of titanium \cite{transitioncooling}. The measured isotope shifts on this line thereby provide key information for future efforts to laser-cool the three bosonic isotopes of titanium.

%Using King analysis \cite{king1,king2,titaniumshifts,iron,hfs_is_uv,ultracoldiron}, these IS can be used to calculate the contributions to the isotope shifts due to uncorrelated electronic motion, correlated electronic motion, and changes in nuclear charge density, thereby providing insight into the atomic and nuclear physics of titanium. Additionally, the isotopic line shifts can be used to calculate the isotope level shifts and relative electron density at the nucleus for all ten levels in the two manifolds. Recent work has shown that the relative electron density at the nucleus can be a powerful benchmark for atomic structure calculations, as the difficult to calculate electron correlation energy is empirically related with this density. \cite{sms,sms2,king1,king2,titaniumshifts,titaniumshifts2,titaniumshifts3,titaniumshifts4}.

%\subsection*{II. Experimental Setup}

The a$^5$F term lies at an energy of about 6800 cm$^{-1}$ above the ground a$^3$F term of titanium. In order to generate a population of metastable atoms, we use a hollow cathode lamp (HCL) (see Refs. \cite{strontium,iron2} for other recent examples of HCL-based atomic spectroscopy), where titanium atoms are sputtered off a titanium cathode bombarded by Ne$^+$ ions. This titanium vapor contains a sufficient steady-state population of atoms in the a$^5$F levels to yield an optical density of around $10^{-3}$ for light resonant with the strongest transitions probed, namely the a$^5$F$_5\!\rightarrow$y$^5$G$_6^\circ$ transition of the most abundant isotope, $^{48}$Ti.

%The optical densities for the other transitions examined in this work were lower by factors that accord with the known variation in line strength with the $J$ and $J^\prime$ values of the lower- and higher-energy states, respectively, and the natural relative abundances of titanium isotopes.

\begin{figure}[t]
    \centering
    \includegraphics[width=0.9 \columnwidth]{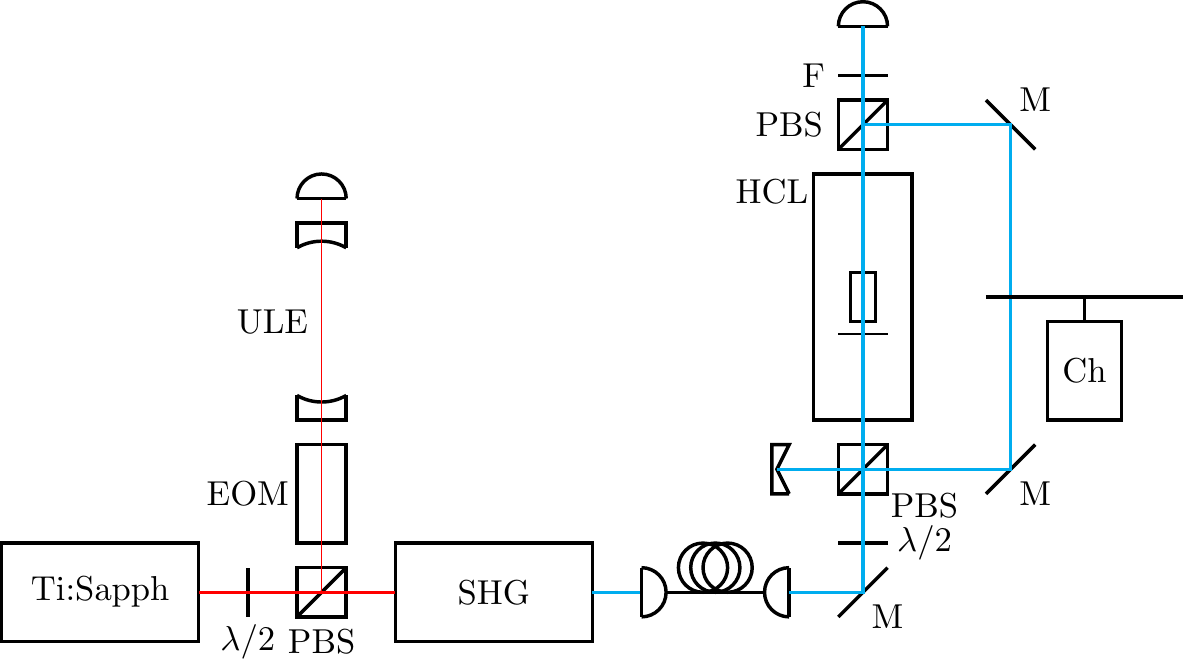}
    \caption{The optical system. Infrared light is first generated with the Ti:Sapph. Part of the light is sent through an EOM to a ULE for frequency referencing, and the rest of it is sent to the SHG stage. The frequency-doubled light is split into pump and probe beams, which counter-propagate through a hollow cathode lamp. A chopper (Ch) amplitude modulates the pump beam. A filtered photodiode detects the probe beam. Key: M, Mirror; $\lambda/2$, half-wave plate; PBS, polarizing beam splitter; F, filter.
}
    \label{fig:optics}
\end{figure}

\begin{figure}[t]
    \includegraphics[width=0.9\columnwidth]{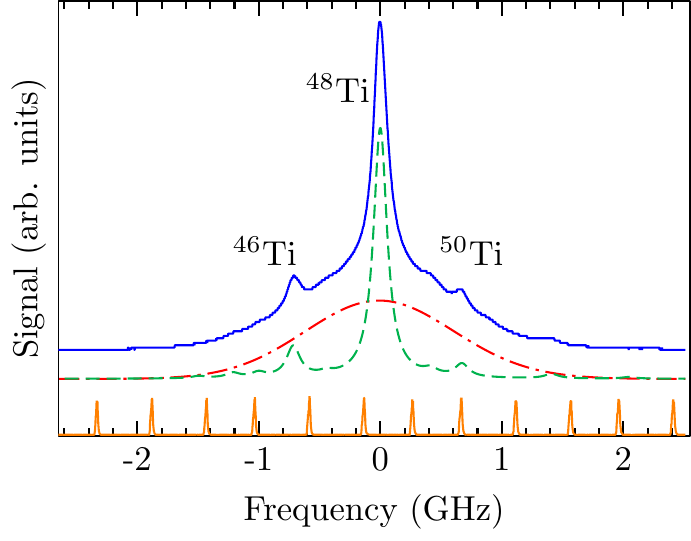}
    \caption{Saturated absorption spectrum of the a$^5$F$_5\!\rightarrow $y$^5$G$^\circ_6$ transition in titanium. The signal (blue, upper) obtained following demodulation at the chopping frequency shows clear peaks for the three $I=0$ isotopes: $^{46}$Ti, $^{48}$Ti, and $^{50}$Ti. The peak frequencies are obtained by fitting the observed spectrum as ten Lorentzians (green, dashed) on top of the Gaussian background (red, dot-dashed) that is produced by diffusion of the excited-state atoms (see text). The number of Lorentzian peaks used in the fitting was chosen to capture as much of the fermionic resonances as detectable without over-constraining the model. ULE transmission resonances (orange, bottom), obtained simultaneously using the electro-optically modulated fundamental laser light, generate calibration frequency markers for the frequency-doubled light used in spectroscopy.}
    \label{fig:chop}
\end{figure}

Laser spectroscopy was performed using the setup presented in Fig. \ref{fig:optics}. A Ti:Sapphire laser \footnote{M-Squared Lasers SolsTiS} (Ti:Sapph), whose emission wavelength was tuned between 996 and 1006 nm, was coupled to a resonant-cavity second harmonic generator (SHG). The fundamental laser frequency was scanned by actuating the length of the Ti:Sapph cavity. The scanned laser frequency was calibrated by sampling the infrared fundamental light and detecting its transmission through the TEM$_{00}$ modes of an ultra-low expansion (ULE) optical cavity \footnote{Stable Laser Systems, 100 kHz linewidth, 1.5 GHz free spectral range}. The ULE cavity provides only sparse frequency markers spaced by the 1.5 GHz free spectral range of the cavity. Additional frequency markers were obtained by adding fixed-frequency sidebands to the light using a broadband electro-optic modulator, resulting in a denser calibration scale for the light, as seen in Fig. \ref{fig:chop}.

The Doppler-broadened single-beam absorption line width of about 1.4 GHz observed in the HCL on the transitions of interest is larger than the isotope shifts we seek to measure. In order to resolve these shifts, we employ saturated absorption spectroscopy. Light from the SHG stage is sent to our saturated-absorption spectroscopy setup, where the beam is split into a pump with intensity near the saturation intensity $I_\mathrm{sat}$ (between 0.1-15 mW/mm$^{2}$ depending on the transition) and a weak (around $I_\mathrm{sat}/50$) probe beam. The absorption signal, which is modulated by chopping the pump beam at 6 kHz using an optical chopper wheel, is then demodulated using a lock-in detector, allowing us to extract a clear Doppler-free absorption signal.

%The probe beam travels through the HCL and transmits the second PBS before reaching a filtered photodetector. The pump beam, amplitude modulated at 6 kHz using an optical chopper, is reflected at each PBS, counter-propagating through the HCL against the probe beam. In-phase demoduation of the output of the photodetector at the chopping frequency using a lock-in amplifier allows for the detection of only the Doppler-free features.

%Titanium's low vapor pressure \cite{pressure} necessitates significant heating to reach optical depths sufficient for precision spectroscopy, especially out of a metastable state. For this reason, a hollow cathode lamp (HCL) is a particularly apt means of producing an atomic sample for spectroscopy \cite{strontium,iron2}. The HCL sustains a population of metastable titanium by bombarding a titanium cathode with a steady stream of Ne$^+$ ions. This comes at the cost of a 1.4 GHz wide Doppler-broadened absorption feature, too broad to resolve different isotopic resonances. To address this, we use saturated absorption spectroscopy (SAS), which provides sub-Doppler absorption signals by using counter-propagating pump and probe beams that selectively address atoms at rest along the beam axis.

%\subsection*{III. Results}
In a typical spectrum (Figure~\ref{fig:chop}), three peaks are evident, corresponding to the three naturally abundant bosonic isotopes. The relative peak amplitudes qualitatively follow the expected isotopic abundance (Table \ref{tab:isotopes}). The measured isotope shifts for these isotopes are presented in Table~\ref{tab:is} in the appendix.

Since the two naturally abundant fermionic isotopes have non-zero nuclear spin and low natural abundance, their already weak resonances are further split between several hyper-fine structure transitions \cite{abundance1,abundance2,hfs_is_uv,titaniumbosons}. Additionally, the observed Doppler-free linewidths in the lamp are broader than the natural linewidths by approximately 60 MHz, arising primarily due to collisions between the metastable titanium and buffer gas in the HCL, which shortens the lifetime of the excited states \cite{collisions,strontium}. Atomic resonances in an HCL are known to suffer from pressure broadening and pressure shifts. Research suggests \cite{pressureSyst1,pressureSyst2,pressureSyst3,pressureSyst4} that pressure shifts are smaller than pressure broadening and that systematic changes to the isotopic shifts, which are differential in nature, are well below the sensitivity of our measurement. This, in addition to the fact that our fitting model deviates most strongly from the experimental data in regions where fermionic peaks would be expected, cause us to suspect that trace absorption peaks from the fermions are present but outside the sensitivity of our measurement.

An additional feature, consisting of a broader Gaussian background superimposed on the atomic resonances, is evident. This feature is likely due to diffusion of excited-state atoms, consistent with theoretical models \cite{collisions} that suggest velocity-changing collisions between the buffer gas and the excited titanium atoms destroy the Doppler-free character of the saturated-absorption measurement scheme for the atoms that undergo collisions. In our system, the mean excited state lifetime of 15 ns is large enough compared to the mean time between collisions of about 150 ns that there is a non-negligible probability that an atom will fall out of the Doppler-free scheme and contribute to this feature.

By fitting the spectra with a physically motivated model that includes all of these features, we can extract the locations of each of the bosonic resonance features, from which we can calculate the isotope shifts. 

The isotope shifts (between isotopes $i$ and $j$) $\mathrm{d}\nu_{ij}$ for a transition can also be decomposed into a normal mass shift $\mathrm{d}\nu_{ij}^{\mathrm{NMS}}$, a specific mass shift $\mathrm{d}\nu_{ij}^{\mathrm{SMS}}$, and a field shift $\mathrm{d}\nu_{ij}^\mathrm{FS}$ \cite{king1,titaniumshifts,iron,collinear,sms,sms2,hfs_is_uv} as 
\begin{equation} \label{eq:decomp}
    \mathrm{d}\nu_{ij} = \mathrm{d}\nu_{ij}^{\mathrm{NMS}} +\mathrm{d}\nu_{ij}^{\mathrm{SMS}} + \mathrm{d}\nu_{ij}^\mathrm{FS}.
\end{equation}

\begin{figure}[!t]
\centering
    \includegraphics[width=0.9\columnwidth]{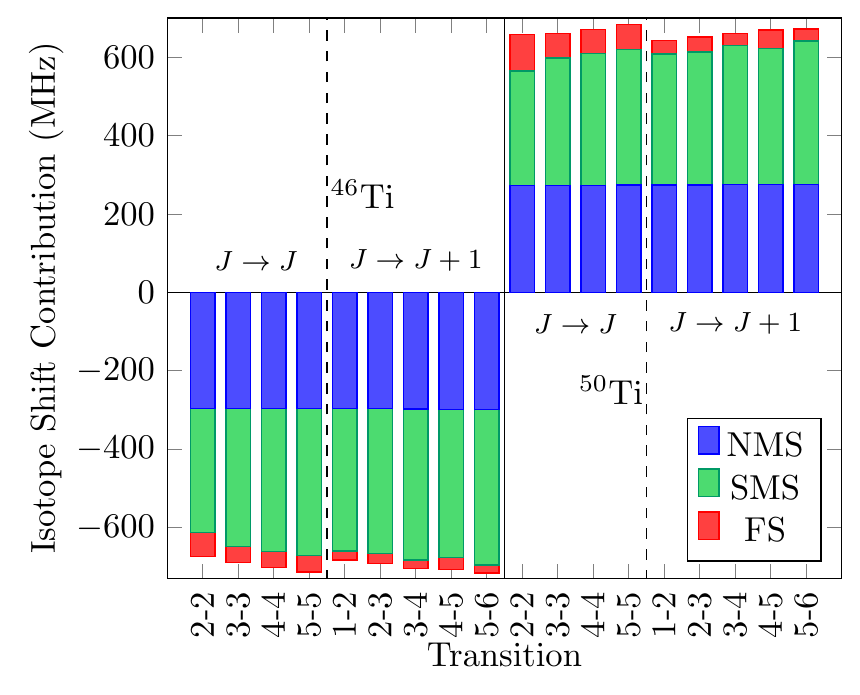}
    \caption{The contributions to the overall isotope shifts due to the normal mass shift (NMS, blue), specific mass shift (SMS, green), and field shift (FS, red) for $^{46}$Ti and $^{50}$Ti, relative to $^{48}$Ti. The contributions sum to the total isotope shift. The horizontal axis labels are the ($J$-$J'$) pairs that define the transitions.}
    \label{fig:contrib}
\end{figure}

\begin{figure*}[!t]
    \centering
    \includegraphics[width=\linewidth]{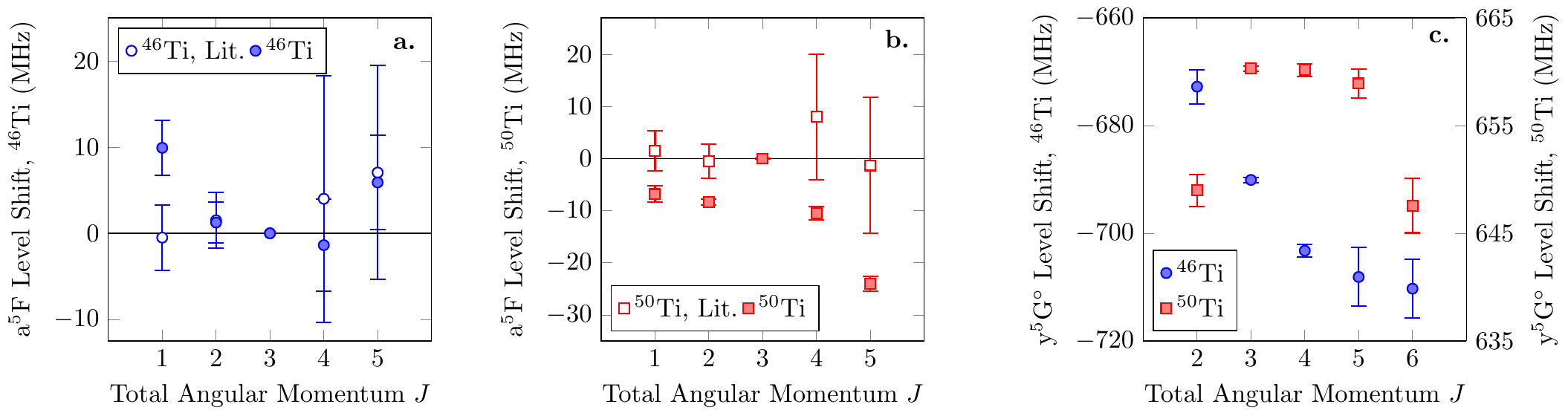}
    \caption{Isotopic level shifts of the a$^5$F manifold (panels a. and b.) and the y$^5$G$^\circ$ manifold (panel c.) for $^{46}$Ti (blue circles) and $^{50}$Ti (red squares), relative to the a$^5$F$_3$ level. Literature values \cite{titaniumshifts} (open symbols) are co-plotted where known. }
    \label{fig:levelshifts}
\end{figure*}{}

\begin{figure}[!h]
 \centering
 \includegraphics[width=0.9 \columnwidth]{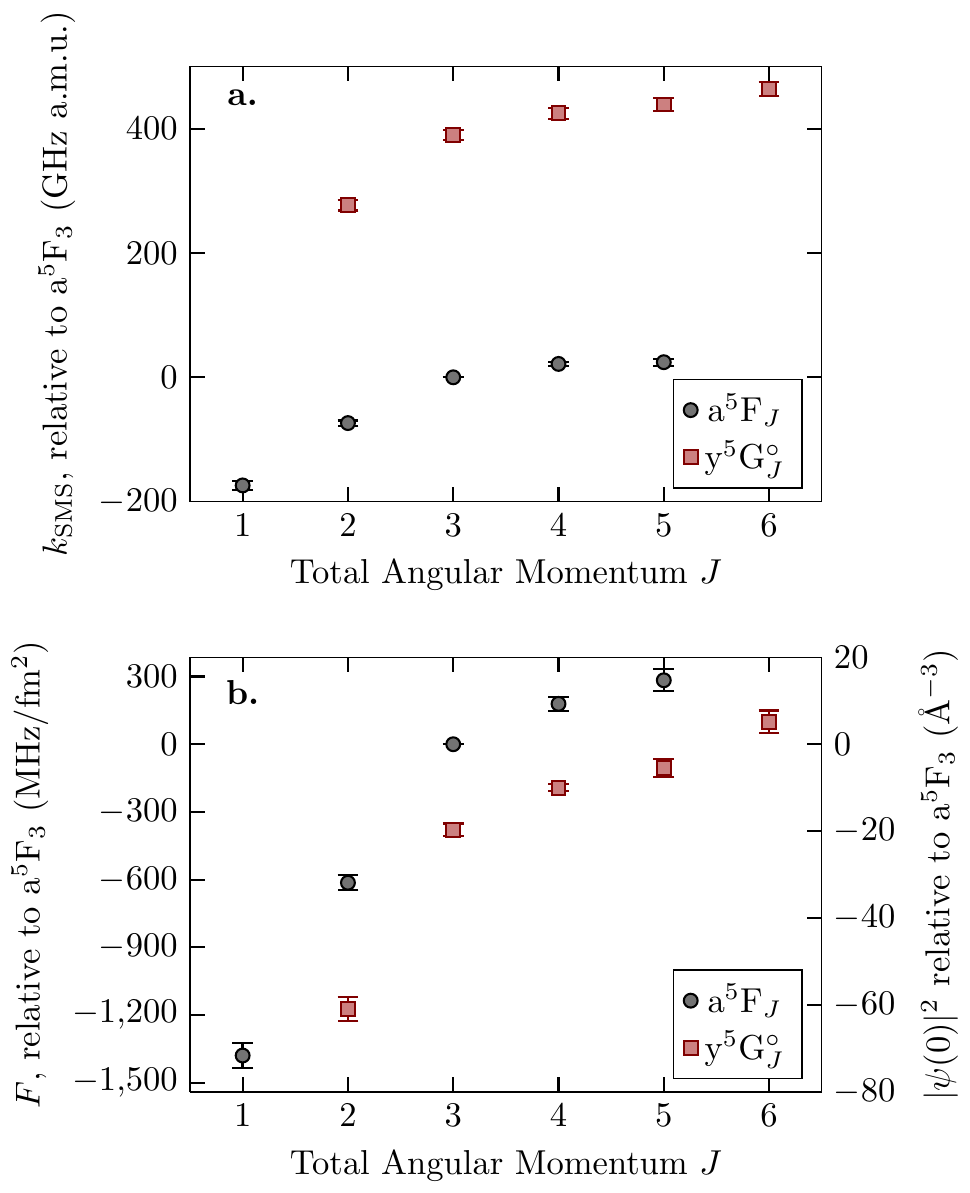}
    \caption{The (a.) specific mass shift and (b.) field shift coefficients for the ten levels investigated, reported relative to a$^5$F$_3$. The value of $F$ is proportional to the electron density at the nucleus (b., right axis).}
    \label{fig:density}
\end{figure}

 The normal mass shift is due to the change in the reduced mass of the system as seen by individual electrons, whereas the specific mass shift is due to change in the correlated motion of electrons. The field shift is due to the difference in charge density of the nucleus between the isotopes. The normal mass shift is readily calculated, and the specific mass and field shifts can be written in terms of specific mass shift and field shift coefficients $k_{\mathrm{SMS}}$ and $F$ \cite{king1,titaniumshifts,king2,iron,collinear,sms,sms2}, respectively, such that equation (\ref{eq:decomp}) reads as
\begin{equation}
      \mathrm{d}\nu_{ij} = (k_{\mathrm{SMS}} + \nu m_e)\frac{m_i-m_j}{m_im_j} + F \, \mathrm{d}\langle r^2\rangle_{ij}
\end{equation}
for isotopic masses $m_i$ and $m_j$, electron mass $m_e$ and difference in mean square nuclear charge radii $\mathrm{d}\langle r^2\rangle_{ij}$. Defining $\mu_{ij} = m_im_j/(m_i-m_j)$ such that
\begin{equation}
    \mu_{ij}\mathrm{d}\nu_{ij} - \nu m_e = k_{\mathrm{SMS}} + F\mu_{ij} \, \mathrm{d}\langle r^2\rangle_{ij},
\end{equation}
we can determine $k_{\mathrm{SMS}}$ and $F$, given our experimental data and $\mathrm{d}\langle r^2\rangle_{ij}$, which are available for titanium from muonic X-ray transition \cite{muonic} measurements in neutral titanium and collinear laser spectroscopy \cite{collinear} measurements in the Ti$^+$ ion. The specific mass shift and field shift coefficients calculated from our data are presented in Table \ref{table:shifts}. %The sign of the field shift is consistent with transitions where the electron density at the nucleus is greater in the lower state than the upper state. This should be the case for these transitions, where one of the 4s electrons in a$^5$F is promoted to the  4p orbital in y$^5$G$^\circ$. % We trace further trends in $J$ to electron shielding effects. 

The values of $k_{SMS}$ and $F$ can be used to calculate the contributions of the specific mass shift and the field shift to the total isotopic line shift. These, along with the calculated normal mass shift, are presented together in Table \ref{tab:kingshifts} and Figure \ref{fig:contrib}. 
The majority of the fractional error in these values comes from the several-percent-level uncertainty in the available literature values of $\mathrm{d}\langle r^2\rangle_{ij}$ \cite{muonic, collinear}.

The specific mass shift is larger than the normal mass shift in all the cases examined in this work, as expected for an atom like titanium with many valence electrons \cite{sms,sms2,hfs_is_uv,tij}. The field shift is weak compared to the mass shifts. Since $F$ is negative, and since $\langle r^2 \rangle$ decreases with the addition of neutrons towards the cusp at the magic $^{50}$Ti nucleus \cite{nuclear,nuclear2,hfs_is_uv}, the field shift contributes in the same direction as the mass shifts.

Since the isotope shift is actually a difference of isotopic level shifts, these isotope shift measurements can also be used to calculate all the level shifts for the a$^5$F and y$^5$G$^\circ$ manifolds, relative to some level, provided the isotope shift on enough lines is measured. By measuring the isotope shift on all five $J\to J+1$ transitions and all four $J\to J$ transitions, we construct a ``ladder'' structure (Figure \ref{fig:levels}) in which we can, by defining one level to have zero isotopic level shift, determine the level shifts of all of the other levels in the two manifolds. We select the a$^5$F$_3$ level as our reference level, since its central location in the a$^5$F manifold allows us to minimize compounding uncertainty. The level shifts are reported, along with literature values calculated using a similar three-manifold ``ladder'' \cite{titaniumshifts} where known, in Table \ref{table:shifts} and Figure \ref{fig:levelshifts}. The level shifts reported here improve upon the precision of literature values by a notable margin and agree to within a 95\% confidence interval. 

A similar ladder scheme can be used to calculate the specific mass and field shift coefficients on each level in the two manifolds. From this, the electron density at the nucleus (relative to the a$^5$F$_3$ level) can be evaluated \cite{king1,density1,density2,density3}.

The values obtained from this analysis, shown in Fig.\ \ref{fig:density}, represent the most significant findings in our work:  We observe a clear, monotonic dependence with $J$ in both the specific mass shift and the field shift, for both the metastable a$^5$F and y$^5$G$^\circ$ levels.  The variation in $k_{SMS}$ reflects a $J$-dependent change in the correlated motion of electrons.  It is compelling to compare the specific mass shift determined from our measurement with theoretically derived electronic wavefunctions for atomic titanium.  

Variation in the field shift is more straightforwardly related to a specific physical quantity, namely the electron density at the location of the nucleus.   We observe that excitation from the metastable  a$^5$F$_J$ levels to the excited y$^5$G$^\circ_J$ levels generally lowers the electron density at the nucleus, a trend that is expected since such excitations imply the promotion of a 4s electron into the 4p orbital. Further, in both manifolds, the electron density shows a strong and clear $J$ dependence. In fact, the $J$-dependent variation of the electron density within each term is far greater than the average difference (about 20 \AA$^{-3}$) between the two terms. This comparison demonstrates the strong electron-correlation effects within the partially filled shells of the titanium atom, and their consequent influence on electronic shielding of the nuclear Coulomb potential.

%\subsection*{IV. Conclusion}
In conclusion, we measure and report the isotope shifts in the nine $\mathrm{a^5F_J\to y^5G^\circ_{J,J+1}}$ transitions in bosonic titanium by performing saturated absorption spectroscopy in a hollow cathode lamp. We use these data to determine the normal mass shift, specific mass shift, and field shift for each stable bosonic isotope and each line. We additionally calculate the isotopic level shifts of all ten levels in the a$^5$F and y$^5$G$^\circ$ energy terms, relative to the a$^5$F$_3$ state. We observe a clear and monotonic variation with $J$ both in the specific mass shift and also in the electron density at the nucleus, for both energy terms examined in this work.  As such, our work provides a strong complement to prior studies of isotope shifts in atomic fine structure, where the atoms and levels examined showed a more complicated $J$ dependence.  Our measurements provide strong benchmarks for theoretical examinations of atomic structure in transition metals.  Additionally, by characterizing isotope shifts in transitions suited for laser cooling of atomic titanium, we open the door to studies of ultracold titanium gases. %These results yield insight into the electronic structure of titanium and are an important first step towards laser cooling and trapping of titanium. The clear $J$-dependent trend in electron density at the nucleus further create opportunities to study the atomic structure of titanium. Using this information, an ultracold atomic ensemble of $^{46}$Ti, $^{48}$Ti, and $^{50}$Ti can be investigated, opening up the possibility of working with novel atomic systems with unique properties.

%\subsection*{Acknowledgements}
We thank Diego Pe\~na, Miguel Aguirre, Diego Novoa, Harvey Hu, Samuel Huber, Johannes Zeiher, Justin Gerber, Emma Deist, and Josh Isaacs for many helpful discussions over the course of this project. We acknowledge support from the Heising-Simons Foundation, from the ARO (contract numbers W911NF1910017 and W911NF2010266, and through the MURI program with grant number W911NF-17-1-0323), and from the California Institute for Quantum Entanglement supported by the Multicampus Research Programs and Initiative of the UC Office of the President (Grant No.\ MRP-19-601445).

\bibliography{ref2.bib}% Produces the bibliography via BibTeX.

\newpage
\appendix

\section{}

In this appendix, we tabulate the numerical data that is discussed in the main body of the paper. Table \ref{tab:isotopes} lists the stable isotopes of titanium and their nuclear spins. Table \ref{tab:is} lists the measured value of the isotope shifts for all of the measured transitions, as well as the King parameters. It should be noted that we have provided statistical and systematic error bars for the quantities listed except for one large systematic error due to the uncertainty in measurements of the radii of titanium's nucleus, which leads to a large uncertainty in the field shift coefficients $F$ and any quantities derived from it. A more precise measurement of d$\langle r^2 \rangle$ would greatly reduce our systematic uncertainties. Table \ref{tab:kingshifts} lists the breakdown of each transition into the normal mass shift (NMS), specific mass shift (SMS) and the field shift (FS), as shown in Figure \ref{fig:contrib}. The systematic uncertainty in the nuclear radius also applies to the breakdown between the SMS and FS, we only list the statistical uncertainty of our measurement. Table \ref{table:shifts} lists the measured isotope shift of each atomic level relative to the a$^5$F$_3$ level, as well as the literature values of these level shifts, where available \cite{titaniumshifts}.

\begin{table*}[]
    \centering
    \begin{tabular}{ c | c c}
        Isotope & Nuclear Spin & Abundance \\\hline
        $^{46}$Ti & 0 & 8.25\%\\
        $^{47}$Ti & 5/2 & 7.44\%\\
        $^{48}$Ti & 0 & 73.72\%\\
        $^{49}$Ti & 7/2 & 5.41\%\\
        $^{50}$Ti & 0 & 5.18\%\\\hline
    \end{tabular}
    \caption{The five naturally abundant isotopes of titanium \cite{abundance1,abundance2}. The three bosonic isotopes ($^{46}$Ti, $^{48}$Ti, and $^{50}$Ti) all have zero nuclear spin, and therefore no hyperfine structure. All five isotopes have more than 5\% natural abundance.
    }
    \label{tab:isotopes}
\end{table*}

\begin{table*}[]
    \centering
    \begin{tabular}{c c | c c |c c}
        Lower & Upper & $^{46}$Ti Shift (MHz) & $^{50}$Ti Shift (MHz) & $k_\mathrm{SMS}$ (GHz a.m.u.) & $F$ (MHz/fm$^2$)\\\hline
        a$^5$F$_1$ & y$^5$G$^\circ_2$ & -682.69(31) & 642.14(38) & 401.36(18)(230) & -206.38(14) \\
        a$^5$F$_2$ & y$^5$G$^\circ_2$ & -674.0 (22) & 657.3 (14) & 350.3(12)(63) & -559.6(21) \\
        a$^5$F$_2$ & y$^5$G$^\circ_3$ & -691.3 (23) & 651.96(50) & 407.5(14)(22) & -234.74(80) \\
        a$^5$F$_3$ & y$^5$G$^\circ_3$ & -690.09(47) & 660.29(28) & 389.29(27)(416) & -379.33(30) \\
        a$^5$F$_3$ & y$^5$G$^\circ_4$ & -703.2 (12) & 660.15(62) & 425.72(72)(202) & -192.46(37) \\
        a$^5$F$_4$ & y$^5$G$^\circ_4$ & -701.9 (52) & 670.6 (11) & 403.2(30)(39) & -372.6(28) \\
        a$^5$F$_4$ & y$^5$G$^\circ_5$ & -706.74(88) & 669.36(39) & 417.15(52)(354) & -283.83(39) \\
        a$^5$F$_5$ & y$^5$G$^\circ_5$ & -714.02(56) & 682.91(56) & 414.37(33)(225) & -388.33(44) \\
        a$^5$F$_5$ & y$^5$G$^\circ_6$ & -716.19(39) & 671.6 (21) & 438.77(25)(202) & -185.89(58) \\\hline
    \end{tabular}
    \caption{The isotope shifts, specific mass shift coefficients, and field shift coefficients for the nine strongest transitions between a$^5$F and y$^5$G$^\circ$. There is a systematic uncertainty in the field and mass shift coefficients arising from uncertainty in literature values \cite{collinear,muonic} of d$\langle r^2\rangle_{ij}$, which have several percent standard fractional error. This manifests as a constant 7.37\% fractional error on the value of $F$ which propagates into the error on $k_\mathrm{SMS}$. For $k_\mathrm{SMS}$, the statistical error is reflected in the first of the two error bounds, and this systematic error is reflected in the second, and for $F$, only the statistical error is reported. Shifts are reported as detunings from the $^{48}$Ti resonance feature.}
    \label{tab:is}
\end{table*}

\begin{table*}[]
\centering
\begin{tabular}{c c | c c c | c c c}
    Lower & Upper & $^{46}$Ti d$\nu^\mathrm{NMS}$ & $^{46}$Ti d$\nu^\mathrm{SMS}$  & $^{46}$Ti d$\nu^\mathrm{FS}$  & $^{50}$Ti d$\nu^\mathrm{NMS}$ & $^{50}$Ti d$\nu^\mathrm{SMS}$  & $^{50}$Ti d$\nu^\mathrm{FS}$ \\\hline
     a$^5$F$_1$ & y$^5$G$^\circ_2$ & -297.0 & -363.5(21) & -22.3(21) & 273.3 & 334.7(30) & 34.1(30)\\
     a$^5$F$_2$ & y$^5$G$^\circ_2$ & -296.4 & -317.3(58) & -60.4(58) & 272.7 & 292.1(85) & 92.3(85)\\
     a$^5$F$_2$ & y$^5$G$^\circ_3$ & -297.4 & -369.1(23) & -25.4(23) & 273.6 & 339.8(34) & 38.7(34)\\
     a$^5$F$_3$ & y$^5$G$^\circ_3$ & -296.5 & -352.7(38) & -41.0(38) & 272.8 & 324.8(55) & 62.6(55)\\
     a$^5$F$_3$ & y$^5$G$^\circ_4$ & -297.9 & -385.5(19) & -20.8(19) & 274.1 & 355.0(28) & 31.8(28)\\
     a$^5$F$_4$ & y$^5$G$^\circ_4$ & -296.7 & -365.1(44) & -40.1(44) & 272.9 & 336.2(65) & 61.3(65)\\
     a$^5$F$_4$ & y$^5$G$^\circ_5$ & -298.4 & -377.8(32) & -30.6(32) & 274.5 & 347.8(47) & 46.8(47)\\
     a$^5$F$_5$ & y$^5$G$^\circ_5$ & -296.9 & -375.2(21) & -41.9(21) & 273.1 & 345.5(30) & 64.1(30)\\
     a$^5$F$_5$ & y$^5$G$^\circ_6$ & -299.0 & -397.4(18) & -20.0(18) & 275.0 & 365.9(27) & 30.5(27)\\\hline
\end{tabular}
\caption{The components of the isotope shifts of bosonic titanium, relative to $^{48}$Ti. All shifts in MHz.}
\label{tab:kingshifts}
\end{table*}

\begin{table*}[]
\centering
\begin{tabular}{l|cc|cc}
Level & Level Shift, $^{46}$Ti & Level Shift, $^{50}$Ti & Literature Level Shift, $^{46}$Ti & Literature Level Shift, $^{50}$Ti\\\hline
%a$^5$F$_1$ & 9.9(32) & -6.8(16) & 4.2(95) & -10.1(74) \\
%a$^5$F$_2$ & 1.3(23) & -8.33(57) & 1.8(72) & -8.2(56)\\
%a$^5$F$_3$ & 0. & 0.& 0.0(59) & 0.0(75) \\
%a$^5$F$_4$ & -1.4(54) & -10.5(12) & 4.5(93) & -8.7(73)\\
%a$^5$F$_5$ & 5.9(55) & -24.0(14) & 22.6(81) & -38.7(104)\\\hline
%y$^5$G$^\circ_2$ & -672.7(32) & 649.0(15) & & \\
%y$^5$G$^\circ_3$ & -690.09(47) & 660.29(28) & & \\
%y$^5$G$^\circ_4$ & -703.2(12) & 660.15(62) & & \\
%y$^5$G$^\circ_5$ & -708.1(54) & 658.9(13) & & \\
%y$^5$G$^\circ_6$ & -710.3(55) & 647.6(25)\\\hline
a$^5$F$_1$ & 9.9(32) & -6.8(16) & -0.5(38) & 1.5(38) \\
a$^5$F$_2$ & 1.3(23) & -8.33(57) & 1.5(32) & -0.5(33)\\
a$^5$F$_3$ & 0. & 0.& 0. & 0. \\
a$^5$F$_4$ & -1.4(49) & -10.5(12) & 4(14) & 8(12)\\
a$^5$F$_5$ & 5.9(55) & -24.0(14) & 7(12) & -1(13)\\\hline
y$^5$G$^\circ_2$ & -672.7(32) & 649.0(15) & & \\
y$^5$G$^\circ_3$ & -690.09(47) & 660.29(28) & & \\
y$^5$G$^\circ_4$ & -703.2(12) & 660.15(62) & & \\
y$^5$G$^\circ_5$ & -708.1(54) & 658.9(13) & & \\
y$^5$G$^\circ_6$ & -710.3(55) & 647.6(25)\\\hline
\end{tabular}
\caption{The experimental and literature \cite{titaniumshifts} (where known) isotopic level shifts of the levels in the a$^5$F and y$^5$G$^\circ$ manifolds, relative to a$^5$F$_3$. All shifts in MHz.}
\label{table:shifts}
\end{table*}

\end{document}